\documentstyle[11pt,psfig,amssymb,amsmath]{article}

\begin{document}

\title{Magnetocapillary Instability of Non--Conducting Liquid Jets 
Revisited : Plateau versus Chandrasekhar}

\author{Yoram Zimmels and Leonid G. Fel\\
\\Department of Civil and Environmental Engineering, \\
Technion, Haifa 32000, Israel}
\maketitle

\def\be{\begin{equation}}
\def\ee{\end{equation}}
\def\p{\prime}  

\begin{abstract}
The magnetostatic and  magnetocapillary instability problems of  
isothermal incompressible and inviscid non--conducting liquid jets in a 
uniform magnetic field, is considered. The equivalence between 
static and dynamic approaches at the onset of instability and cut--off 
wavelength is shown. It is established that in the absence of electric 
currents the stability of permeable jets can be changed by the magnetic 
field. A new dispersion relation for magnetocapillary instability in such 
jets is derived. This relation differs from that given by Chandrasekhar. 
The existence of critical magnetic field which stabilizes jets with 
finite susceptibility is established. It is shown that the jet is 
stabilized by the field irrespective of its being para-- or diamagnetic, 
but the extent of stabilization is different.
\end{abstract}

\vskip .5cm
\centerline{{\bf Key words:} Plateau problem, Magnetocapillary  
instability, Non--conducting liquid jet.}

\vskip 5.5cm
\centerline{e-mail: lfel@techunix.technion.ac.il}

\newpage

\noindent
\section{Introduction}
\label{introd}
Joseph Antoine Ferdinand Plateau (1801--1883) was a Belgian physicist who 
is best remembered in mathematics for Plateau problem. He wrote his seminal 
book \cite{Plat73} being already blind for the last 40 years of his life. 
The Plateau problem, in its brief formulation, is to find a surface of 
minimum area $S$ given its boundary $\partial \Omega$. The problem 
relates to the principle of 
minimum free energy at equilibrium. In this context it may be required  
to find a surface with isotropic tension $\sigma$ which provides a minimum 
surface free energy $\sigma S$. Rayleigh \cite{Rayl79} gave a theoretical 
explanation for the instability of liquid cylinders that are longer than 
their circumference. Further generalization is called for if 
the excess free energy $W$ of the cylinder comprises different types of 
energy that reflect a more complex structure of the liquid (e.g. 
elasticity \cite{fel02} {\it etc}) and well as its capacity to interact 
with external fields. The threshold of static instability, which is 
associated with the change of sign of $W$, when the disturbed 
configuration becomes more preferable, is defined by
\begin{eqnarray}
W(kR,p)=0\;\;\rightarrow\;\;k_s=k_s(p)\;,
\label{intro00}
\end{eqnarray}
where $k$ is a wave number assigned to a small surface disturbance, $R$ is 
a radius of cylinder, and the parameter $p$ stands for the effect of an  
external interaction, which contributes to the jet evolution, e.g. 
fluid polarization in the presence of external fields. Note that the 
{\it static cut--off wave number} $k_s$ is related to the {\it static 
cut--off wavelength} $\Lambda_s$ by $\Lambda_s(p)=2\pi/k_s$.

Plateau instability being a static problem has also a dynamic aspect. Consider 
Rayleigh instability in liquid jets of radius $R$ and its corresponding 
dispersion equation $s=s(kR,p)$. The latter determines the exponential 
evolution in time $\Theta({\bf r},t)\sim e^{st}$ of all hydrodynamic 
functions $\Theta({\bf r},t)$ of the jet, with the growth rate $s$. Rayleigh's 
theory of capillary instability in liquid jets states that the maximum of 
the dispersion function $s(kR,p)$ which corresponds to the wave number 
$k_{max}(p)$, gives rise to evolution of the largest capillary instability. 
The range of wave numbers $k$ which contribute to the evolution of the 
instability is given by $s>0$. Thus, the threshold of instability follows as
\begin{eqnarray}
s(kR,p)=0\;\;\rightarrow\;\;k_d=k_d(p)\;.
\label{intro0}
\end{eqnarray}
Note that the {\it dynamic cut--off wave number} $k_d$ is related to the 
{\it dynamic cut--off wavelength} $\Lambda_d$ by $\Lambda_d(p)=2\pi/k_d$.
The value $\Lambda_d(0)$, which corresponds to free jet evolution is equal 
$2\pi R$ according to Rayleigh \cite{Rayl79}. If there exist a critical 
parameter $p=p_{cr}$ such that for all $p\geq p_{cr}$ the expression 
$s(kR,p)\leq 0$ holds, then for $p\geq p_{cr}$ the jet is stable for all 
wavelengths. The latter means that the liquid jet preserves its initial 
shape which is unaffected by small perturbations irrespective to the jet 
velocity. Thus, this conclusion must hold also in the limiting case of 
motionless fluid, i.e.
\begin{eqnarray}
\Lambda_d(p)\equiv \Lambda_s(p)\;.
\label{intro10}
\end{eqnarray}
This identity reflects a deep equivalence between the static 
approach (the threshold of static instability concerned with an excess 
free energy) and the dynamic approach (the bifurcation of the first 
non--trivial steady state of the inviscid hydrodynamic system). 
Thus, we come to Plateau problem for liquid cylinder with free surface 
in an external field.

The capillary instability of magnetizable liquid jets in the presence of 
an axial magnetic field constitutes a classical example. Chandrasekhar 
\cite{Chan61} treated the hydrodynamics of magnetic liquid jets and 
derived their dispersion relation. This relation reads particularly 
simple for superconducting $s_{sc}(\varpi)$ and non--conducting 
$s_{nc}(\varpi)$  liquids (see \cite{Chan61}, p. 545, formula (165) and p. 
549, formula (205))
\begin{eqnarray}
s_{sc}^2(\varpi)=\frac{\sigma}{R^3 \rho}
\left\{\frac{\varpi I_1(\varpi)}{I_0(\varpi)}(1-\varpi^2)-
\left(\frac{H}{H_{s}}\right)^2
\frac{\varpi }{I_0(\varpi)K_1(\varpi)}\right\}\;,\;\;\;
s_{nc}^2(\varpi)=\frac{\sigma}{R^3 \rho} \frac{\varpi 
I_1(\varpi)}{I_0(\varpi)}(1-\varpi^2)\;,
\label{introd01}
\end{eqnarray}
where $\varpi=kR$ and $H_{s}=\sqrt{\sigma/\mu_0(1+\chi) R}$ is designated  
by Chandrasekhar as characteristic field. $\chi$, $\sigma,\rho$ stand for 
magnetic susceptibility, isotropic surface tension and density of the 
liquid, respectively, and $\mu_0$ denotes the permeability of free space. 
$I_m(x)$ and $K_m(x)$ are the modified Bessel functions of order $m$  
of the 1st and 2nd kind respectively. In the superconducting limit the 
relation (\ref{introd01}) discloses the existence of critical magnetic 
field $H_{cr}=H_{s}/\sqrt{2}$ beyond which the jet is stable. In the 
non--conducting limit, this dispersion relation concides with that 
of Rayleigh \cite{Rayl79} and `{\it in this limit the magnetic field has 
no effect on the capillary instability as should, indeed, be the case}' 
(quotation from \cite{Chan61}, Chap. 12, \S  112, p. 549).

The latter result seems rather unexpected. Indeed, the absence of the electric 
currents in the presence of static magnetic field does not exclude coupling 
between the magnetic dipole moments of the liquid and the external field. 
Since the linear problem sets no constraints on the initial velocity of 
the jet, the above observation must be also correct in statics (Plateau 
problem) when the magnetic cylinder is destabilized. We proceed to verify our 
observation, first for the case of Plateau instability (see Sections 
\ref{energ}, \ref{maxwe}, \ref{instab}) and then move on to the dynamic case. 
In Section \ref{noncond} we give an accurate solution of the magnetocapillary 
instability problem of isothermal, incompressible, inviscid and non--conducting 
jets, in the presence of a uniform magnetic field. We show that the dispersion 
relation differs from the one (\ref{introd01}) found by Chandrasekhar 
\cite{Chan61}. The new relation accounts for the effect of magnetic fields and 
conforms with the solution of Plateau instability for magnetizable liquids. 
The reason for the discreapency in Chandrasekhar treatment arises due to 
incorrect boundary conditions which he applied to the magnetic field
\footnote{We quote from  \cite{Chan61} ( Chap. 12, \S  112, p. 547) : 
` {\it the field {\bf h} must be continuous across the boundary} '. The correct 
boundary conditions require continuity of the tangential component of the 
magnetic field and of the normal component of the magnetic induction (see 
equation (\ref{max2}) in Section \ref{maxwe} and equation (\ref{bonstr10}) in 
Section \ref{boundar1} of the present paper).}.

\section{Free energy of liquid cylinder in the presence of magnetic field.}
\label{energ}
The Plateau problem of static instability of a non--conducting liquid 
cylinder which is subjected  to a uniform magnetic field appears, at 
first glance, to be a simple generalization of its counterpart in the 
absence of external fields. However, deeper consideration shows that 
the presence of the field complicates considerably the physical picture 
and computational procedure. A fundamenatal question arises concerning 
the correct definition of the excess free energy $W$ which must be 
minimized via variation of the cylinder shape.

Consider {\it an isothermal liquid cylinder in a uniform magnetic field} 
${\bf H_0}$ that is applied in free space along the cylinder axis. The 
magnetic susceptibility $\chi$ of the cylinder is assumed isotropic, 
independent of magnetic field,  and satisfying the thermodynamic condition 
$\chi>-1$ \cite{land84}. When the liquid cylinder is undisturbed the 
total free energy $F^0$ of the system is given by
\begin{equation}
F^0={\cal E}_s^0-\frac{\chi\mu_0H_0^2}{2}\cdot 
\int_{\Omega_{cyl}^0}dv\;,
\label{stat1}
\end{equation}
where the integral represents the volume $\pi R^2L$, enclosed by the area
$\partial \Omega_{cyl}^0$ which is occupied by undisturbed cylinder. The 
term ${\cal E}_s^0=\sigma \int_{\partial \Omega_{cyl}^0}ds=2\pi \sigma R L$ 
stands for the surface free energy of the undisturbed cylinder, where $R,L$ 
and $\sigma$ denote its radius, length and surface tension respectively. 
Deformation of the cylinder shape changes the field ${\bf H}({\bf r})$ 
over all space ${\mathbb R}^3$, i.e. in both the internal domain $\Omega_{cyl}$ 
and its complement (the exterior domain) ${\mathbb R}^3\setminus \Omega_{cyl}$. 
Following Plateau, we assume conservation of the cylinder volume 
\begin{equation}
\int_{\Omega_{cyl}^0}dv=\int_{\Omega_{cyl}}dv\;.
\label{stat2a}
\end{equation}
The total free energy $F$ of the disturbed cylinder takes the following form, 
\begin{equation}
F={\cal E}_s-(1+\chi)\frac{\mu_0}{2}\cdot
\int_{\Omega_{cyl}}\;^{\sf in}{\bf H}^2({\bf r})dv- \frac{\mu_0}{2}\cdot
\int_{{\mathbb R}^3\setminus \Omega_{cyl}}\;^{\sf ex}{\bf H}^2({\bf r})dv+
\frac{\mu_0}{2}\cdot \int_{{\mathbb R}^3}H_0^2dv
\label{stat3}
\end{equation}
where $^{\sf in}{\bf H}({\bf r})$ and $^{\sf ex}{\bf H}({\bf r})$ denote the 
internal and external magnetic fields, and ${\cal E}_s=\sigma \int_{\partial 
\Omega_{cyl}}ds$. The excess free energy $W$ of the system is defined as, 
\begin{equation}
W=F-F^0\;.
\label{stat5}
\end{equation}
From the mathematical standpoint, the variational problem for minimization 
of $W$, supplemented with constraint (\ref{stat2a}) for all smooth 
surfaces $\partial \Omega_{cyl}$, is known as the isoperimetric problem.

The cylinder instability can be studied assuming small perturbation in its 
shape. In this case  the Plateau problem becomes solvable in closed form.

Let the extent of deformation be characterized by a length $\zeta_0$, such  
that $\zeta_0/R=\epsilon\ll 1$. Then the following approximation holds
\begin{equation}
\int_{\partial \Omega_{cyl}}ds-\int_{\partial \Omega_{cyl}^0}ds\simeq 
A\epsilon^2RL\;,\;\;\;\;\;\;
\label{stat6} 
\end{equation}
where $A$ is a dimensionless parameter dependent on the deformed geometry.

Here, the fields $^{\sf in}{\bf H}({\bf r})$ and $^{\sf ex}{\bf H}({\bf r})$ 
which must satisfy Maxwell equations can be represented as small 
perturbations of ${\bf H}_0$,
\begin{equation}
^{\sf in}{\bf H}({\bf r})={\bf H}_0+^{\sf in}\!\!{\bf H}^1({\bf r})=
\left(H_0+^{\sf in}\!\!H_z^1,\;^{\sf in}H_r^1\right)\;,\;\;\;
^{\sf ex}{\bf H}({\bf r})={\bf H}_0+^{\sf ex}\!\!{\bf H}^1({\bf r})=
\left(H_0+^{\sf ex}\!\!H_z^1,\;^{\sf ex}H_r^1\right)\;,
\label{stat7}
\end{equation}
where according to the assumption $\epsilon\ll 1$ the following 
approximations apply (see Setion \ref{maxwe})
\begin{equation}  
\left\{^{\sf in}H_r^1,\;^{\sf ex}H_r^1,\;^{\sf in}H_z^1,\;^{\sf 
ex}H_z^1\right\}=
\left\{^{\sf in}h_r^1,\;^{\sf ex}h_r^1,\;^{\sf in}h_z^1,\;^{\sf 
ex}h_z^1\right\}\times \epsilon \chi H_0\;.
\label{stat8}
\end{equation}
The dimensionless fields $^{\sf in,ex}h_{r,z}^1(r,z)$ are dependent on the 
coordinates. Inserting (\ref{stat8}) into (\ref{stat3}), and performing 
the integration, we evaluate $W$ given in (\ref{stat5}) as, 
\begin{eqnarray}
W=A\sigma \epsilon^2RL-\frac{\mu_0}{2}U\;,
\label{stat8a}
\end{eqnarray}
where use was made of (\ref{stat1}) and $U$ is given by (see Appendix 
\ref{appendix1})
\begin{eqnarray}
U&=&(1+\chi)\int_{\Omega_{cyl}}
\left\{\;\left(^{\sf in}H_z^1\right)^2+\left(^{\sf in}H_r^1\right)^2\right\}dv+
\int_{{\mathbb R}^3\setminus \Omega_{cyl}}
\left\{\;\left(^{\sf ex}H_z^1\right)^2+\left(^{\sf ex}H_r^1\right)^2\right\}dv+
\nonumber\\
&&2H_0\left((1+\chi)\int_{\Omega_{cyl}}\;^{\sf in}H_z^1dv+
\int_{{\mathbb R}^3\setminus \Omega_{cyl}}\;^{\sf ex}H_z^1dv\right)\;.
\label{www1}
\end{eqnarray}
We specify the commonly used harmonic deformation of the cylinder as 
$r(z)=R+\zeta_0\cos kz$, where $k=2\pi/\Lambda$, $\Lambda$ being the 
disturbance wavelength. By virtue of translational invariance of the problem
\begin{eqnarray}
^{\sf in,ex}h_{r,z}^1(r,z+\Lambda)=\;^{\sf in,ex}h_{r,z}^1(r,z)
\label{stat9}
\end{eqnarray}
we set $L=\Lambda$ and evaluate the free energy per unit wave length. 
For this type of deformation the parameter $A$ is known \cite{Plat73},
\begin{eqnarray}
\sigma \int_{\partial \Omega_{cyl}}ds-2\pi \sigma R 
L=\sigma\frac{\pi \zeta_0^2}{2 R}L
\left(\varpi^2-1\right)\;\;\;\;\longrightarrow\;\;\;\;
A=\frac{\pi}{2}\left(\varpi^2-1\right)\;,\;\;\;\varpi=kR\;.
\label{stat10}
\end{eqnarray}
Next we proceed to solve the boundary problem and get the distribution of 
the magnetic fields.
\section{Boundary problem and its solution}
\label{maxwe}
The magnetostatics of the disturbed liquid cylinder is governed by Maxwell 
equations for the internal $^{\sf in}{\bf H}({\bf r})$ and external 
$^{\sf ex}{\bf H}({\bf r})$ magnetic fields
\begin{eqnarray}
{\rm rot}\;^{\sf in}{\bf H}={\rm rot}\;^{\sf ex}{\bf H}=0\;,\;\;\;\;
{\rm div}\;^{\sf in}{\bf B}={\rm div}\;^{\sf ex}{\bf B}=0\;,\;
\label{max1}
\end{eqnarray}
where $^{\sf in}{\bf B}=\mu_0(1+\chi){}^{\sf in}{\bf H}$ and 
$^{\sf ex}{\bf B}=\mu_0{}^{\sf ex}{\bf H}$ denote internal and external 
magnetic inductions, respectively. Equations (\ref{max1}) must be 
supplemented with boundary conditions (BC) at the interface $r=R$, 
\begin{eqnarray}
\langle {}^{\sf in}{\bf H},{\bf t}\rangle=
\langle {}^{\sf ex}{\bf H},{\bf t}\rangle\;,
\;\;\;\;
\langle {}^{\sf in}{\bf B},{\bf e}\rangle=
\langle {}^{\sf ex}{\bf B},{\bf e}\rangle\;,\;\;\;\;
\langle {\bf e},{\bf t}\rangle=0\;,
\label{max2}
\end{eqnarray}
where {\bf t} and {\bf e} stand for tangential and normal unit vectors to 
the surface, respectively. Since the surface deformation is small, 
linearization can be applied,
\begin{eqnarray}
t_r=-e_z=\partial \zeta/\partial z\;,\;\;\;t_z=e_r=
\sqrt{1-\left(\partial \zeta/\partial z\right)^2}\simeq 1\;.
\label{max2a}
\end{eqnarray}
A standard way to solve the problem is to introduce the magnetic potentials
$\Phi_{\sf in}({\bf r})$ and $\Phi_{\sf ex}({\bf r})$ which are defined as 
${}^{\sf in}{\bf H}^1({\bf r})=-\nabla\Phi_{\sf in}, 
{}^{\sf ex}{\bf H}^1({\bf r})=-\nabla\Phi_{\sf ex}$, where 
$|\nabla\Phi_{\sf in}|, |\nabla\Phi_{\sf ex}|\ll H_0$. These potentials 
satisfy the first two equations in (\ref{max1}). The last two equations in
(\ref{max1}) yield,
\begin{eqnarray}
\frac{\partial^2 \Phi_{\sf in}}{\partial z^2}+\Delta_2 
\Phi_{\sf in}=0\;,\;\;\;\;
\frac{\partial^2 \Phi_{\sf ex}}{\partial 
z^2}+\Delta_2\Phi_{\sf ex}=0\;,\;\;\;\;\;\;
\Delta_2=\frac{\partial^2 }{\partial r^2}+\frac{1}{r}
\frac{\partial }{\partial r}\;,
\label{max3}
\end{eqnarray}
where $\Delta_2$ is the two--dimensional Laplacian. Reformulation of the 
BC (\ref{max2}) for $\Phi_{\sf ex}({\bf r})$, $\Phi_{\sf in}({\bf r})$ 
gives,
\begin{eqnarray}
\frac{\partial \Phi_{\sf ex}}{\partial z}=
\frac{\partial \Phi_{\sf in}}{\partial z}\;,
\;\;\;\;\;\frac{\partial \Phi_{\sf ex}}{\partial r}-\left(1+\chi\right)   
\frac{\partial \Phi_{\sf in}}{\partial r}=\chi H\frac{\partial 
\zeta}{\partial z}\;.
\label{max4}
\end{eqnarray}
Using  $\Phi_{\sf in}(r,z)=\phi_{\sf in}(r)\sin kz$ and 
$\Phi_{\sf ex}(r,z)=\phi_{\sf ex}(r)\sin kz$ we find
\begin{equation}
\left(\Delta_2-k^2\right)\phi_{\sf ex}=0\;,\;\;\;\;\;
\left(\Delta_2-k^2\right)\phi_{\sf in}=0\;
\label{max5}
\end{equation}
with BC at $r=R$
\begin{equation}
\phi_{\sf ex}=\phi_{\sf in}\;,\;\;\;\;
(1+\chi)\frac{\partial \phi_{\sf in}}{\partial r}-
\frac{\partial \phi_{\sf ex}}{\partial r}=\chi Hk\zeta_0\;.
\label{max6}
\end{equation}
The solutions of (\ref{max5}), which satisfy BC (\ref{max6}) and are 
finite at $r=0$ and $r=\infty$, are obtained as,
\begin{equation}
\phi_{\sf in}(r)=\zeta_0\chi H_0\frac{b(\varpi,\chi)}{\varpi}I_0(kr)\;,\;\;\;
\phi_{\sf ex}(r)=\zeta_0\chi H_0\frac{c(\varpi,\chi)}{\varpi} K_0(kr)\;,
\label{max7}
\end{equation}
where 
\begin{eqnarray}
b(\varpi,\chi)=\varpi^2\frac{K_0(\varpi)}{T(\varpi,\chi)}\;,\;\;\;
c(\varpi,\chi)=\varpi^2\frac{I_0(\varpi)}{T(\varpi,\chi)}\;,\;\;\;
T(\varpi,\chi)=1+\chi \varpi I_1(\varpi)K_0(\varpi)\;.
\label{max7a}
\end{eqnarray}
Recalling the definition (\ref{stat8}) of the dimensionless fields $^{\sf 
in,ex}h_{r,z}^1(r,z)$, the following final expressions are obtained,
\begin{eqnarray}
&&{}^{\sf in}h_z^1=-b(\varpi,\chi)I_0(kr)\cos kz\;,\;\;
{}^{\sf in}h_r^1=-b(\varpi,\chi)I_1(kr)\sin kz\;,
\nonumber\\
&&{}^{\sf ex}h_z^1=-c(\varpi,\chi)K_0(kr)\cos kz\;,\;\;
{}^{\sf ex}h_r^1=c(\varpi,\chi)K_1(kr)\sin kz\;.
\label{max8}
\end{eqnarray}
Bearing in mind that the function $Q(\varpi)=\varpi I_1(\varpi)K_0(\varpi)$ 
is monotone growing, at the positive half axis and is bounded $0\leq Q(\varpi)
< 1/2$, we conclude that the fields $^{\sf in,ex}h_{r,z}^1(r,z)$ 
are free of singularities in the thermodynamically relevant region $\chi>-1$.
\section{Plateau instability}
\label{instab}
In this Section we calculate via equation (\ref{stat8}) the excess free energy 
$W$ in terms of the dimensionless fields ${}^{\sf ex}h_{z,r}^1,{}^
{\sf in}h_{z,r}^1$. Inserting (\ref{max8}) into (\ref{www1}) we get (see 
Appendix \ref{appendix1})
\begin{eqnarray}
W=\frac{\pi L\sigma R}{2}\epsilon^2 \cdot f(\varpi,\chi,H_0)\;,\;\;\;
f(\varpi,\chi,H_0)=\varpi^2-1+\chi^2\varpi^2\frac{\mu_0 R H_0^2}{\sigma}
\frac{I_0(\varpi)K_0(\varpi)}{T(\varpi,\chi)}\;,
\label{fff1}
\end{eqnarray}
where the dimensionless excess free energy $f(\varpi,\chi,H_0)$ can be 
defined by introducing the characteristic field $H_{Ch}$
\begin{eqnarray}
f(\varpi,\chi,H_0)&=&\varpi^2-1+\chi\varpi^2
\frac{I_0(\varpi)K_0(\varpi)}{T(\varpi,\chi)}
\left(\frac{H_0}{H_{Ch}}\right)^2\;,\;\;\;
H_{Ch}=\sqrt{\frac{\sigma}{\mu_0\chi R}}\;.
\label{finl1}
\end{eqnarray}
Notice that $H_{Ch}$ differs from $H_s$ introduced by Chandrasekhar in 
(\ref{introd01}). Formula (\ref{finl1}) serves for both paramagnetic 
($\chi>0$) and diamagnetic ($\chi<0$) liquids, since the latter case does 
not lead to an imaginary expression in (\ref{finl1}) due to 
the term $\chi^2$ in (\ref{fff1}). 

\begin{figure}[h]
\centerline{\psfig{figure=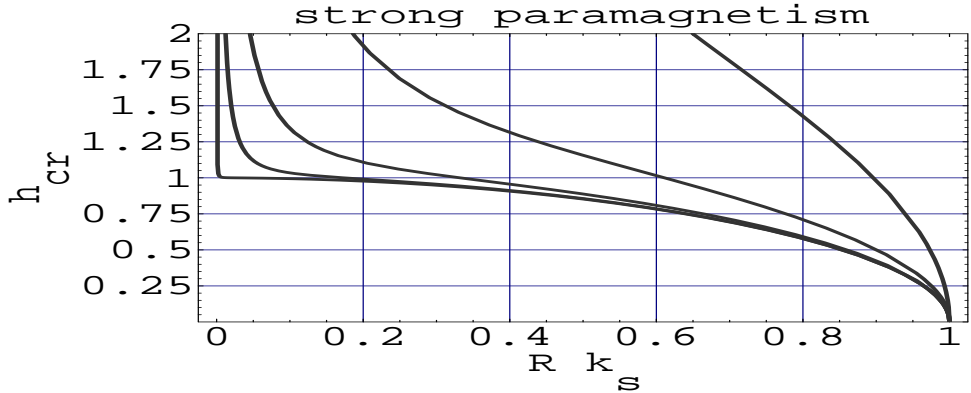,height=5.5cm,width=14cm}}
\vspace{-.5cm}
\caption{A plot of $h_{cr}$ {\it versus} $k_sR$ for large positive 
susceptibilities : $\chi=1,10,10^2,10^3,10^6$, from right to left, 
respectively.}
\label{figplat1}
\end{figure}

\begin{figure}[h]
\centerline{\psfig{figure=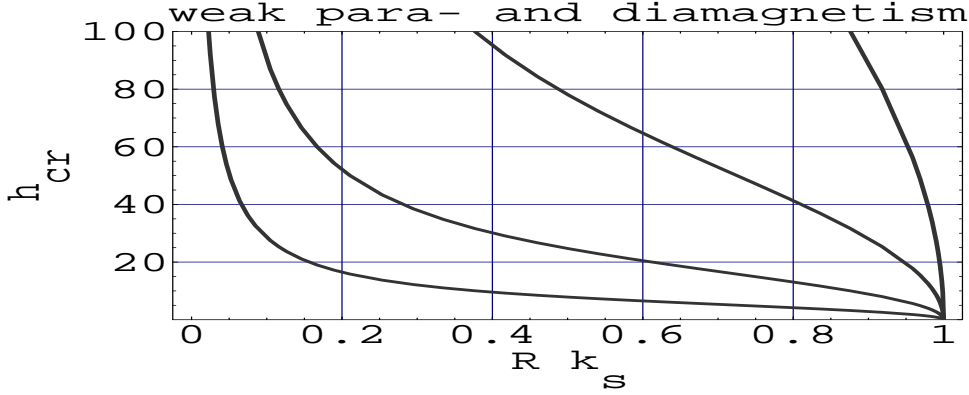,height=5.5cm,width=14cm}}
\vspace{-.5cm}
\caption{A plot of $h_{cr}$ {\it versus} $k_sR$ for small positive and negative 
susceptibilities: $\chi=\pm 10^{-4},\pm 10^{-3},\pm 10^{-2},\pm 10^{-1}$, from 
right to left, respectively. The plots corresponding to diamagnetic and 
paramagnetic cases nearly coincide.}
\label{figplat2}
\end{figure}

Consider the field $H_{cr}(\varpi,\chi)$ satisfying $f(\varpi,\chi,H_{cr})=0$ 
and call it {\it critical field}. Beyond this field, $H_0\geq H_{cr}$, the 
cylinder is stable. The expession for the critical field is
\begin{eqnarray}
H_{cr}(\varpi,\chi)=\frac{H_{Ch}}{\varpi}\sqrt{\frac{1-\varpi^2}{\chi}  
\frac{T(\varpi,\chi)}{I_0(\varpi)K_0(\varpi)}}\;.
\label{critpos1}
\end{eqnarray}
Figures \ref{figplat1}, \ref{figplat2} show plots of $h_{cr}=H_{cr}/H_{Ch}$ 
for strong ($\chi \gg 1$) and weak ($|\chi|\ll 1$) magnetic liquids.

The corresponding asymptotics for $H_{cr}(\varpi,\chi)$, in the case of weak 
$\chi\ll 1$ and strong $\chi\gg 1$ magnetic susceptibilities and in a long 
wave limit $\varpi\to 0$, which are derived in Appendix \ref{appendix2}, 
are presented below 
\begin{eqnarray}  
&&\varpi\to 0\;,\;\;|\chi|\ll 1\;,\;\;\;\;\;\;\;
\frac{H_{cr}(\varpi,\chi)}{H_{Ch}}=\frac{1}{\varpi\sqrt{-\chi\ln\varpi}}
\left(1-\frac{\chi}{2}\varpi^2\ln\varpi\right)\;,\label{critpos2}\\
&&\varpi\to 0\;,\;\;\chi\to \infty\;,\;\;\;\;\;\;\;
\frac{H_{cr}(\varpi,\chi)}{H_{Ch}}=\frac{1}{\sqrt{2}}
\left(1-\frac{9}{16}\varpi^2\right)\;,\label{critpos3}\\
&&\varpi\to 1\;,\;\;\;\;\;\;\;\;\;\;\;\;\;\;\;\;\;\;\;\;\;\;\;
\frac{H_{cr}(\varpi,\chi)}{H_{Ch}}= B_1 \sqrt{B_2+\frac{1}{\chi}}
\sqrt{1-\varpi^2}\;,
\label{critpos4}
\end{eqnarray}
where $B_1=1/\sqrt{I_0(1)K_0(1)}\simeq 1.3697,\;B_2=I_1(1)K_0(1)\simeq 0.2379$.

To the end of this Section we present physical arguments which justify our 
conclusion regarding stabilization of the permeable jets for both paramagnetic 
and diamagnetic liguids. The orientation of the dipole moments along the 
applied magnetic field in the paramagnetic liquid enforces the rigidity of 
the jet under small disturbances. In the case of diamagnetic liquid the 
increase of the jets' stabilization is due to orientation of the dipole moments 
in the opposite direction which also enforces the rigidity of the jet. The 
similar increase of the rigidity, and its corresponding influence on the 
stabilization, was found recently \cite{fel02} in the liquid crystalline jet 
due to elasticity of the media. Although the critical fields for paramagnetic 
$H_{cr}(\varpi,\chi)$ and diamagnetic $H_{cr}(\varpi,-\chi)$ liquids are 
different the following universal relation holds,
\begin{eqnarray}
\frac{H^2_{cr}(\varpi,\chi)}{H_{Ch}^2}-
\frac{H^2_{cr}(\varpi,-\chi)}{H_{Ch}^2}=2\frac{1-\varpi^2}{\varpi}
\frac{I_1(\varpi)}{I_0(\varpi)}\;.
\label{critpos11}
\end{eqnarray}
\section{Hydrodynamics of Non--Conducting Jet in a Magnetic Field}
\label{noncond}
Consider an isothermal, incompressible, inviscid and non--conducting jet 
in the presence of a magnetic field ${\bf H}_0$  applied along its $z$--axis. 
The deviation from initial values of the pressure is defined as $P_1^{\sf 
in}=P^{\sf in}-P^{\sf in}_0$, where $P_0^{\sf in}$ is the unperturbed 
pressure within the cylindrical jet. The deviations of the internal and 
external magnetic fields are defined as $^{\sf in}{\bf H}^1=^{\sf in}\!{\bf H}-
{\bf H}_0$ and $^{\sf ex}{\bf H}^1=^{\sf ex}{\bf H}-{\bf H}_0$, respectively. 
The governing equations of magnetohydrodynamics which are given by
\begin{eqnarray}
&&{\rm div}{\bf V}^{\sf in}=0\;,\;\;\;\;
\rho \frac{\partial V_j^{\sf in}}{\partial t}=-
\frac{\partial T_{jk}^{\sf in}}{\partial x_k}\;,\;\;\;\;
^{\sf in}T_{jk}=\left(P^{\sf in}+
\mu\frac{^{\sf in}{\bf H}^2}{2}\right)\delta_{jk}-
\mu\;^{\sf in}H_j\;^{\sf in}H_k\;,\label{nav1}\\
&&{\rm div}\; ^{\alpha}{\bf H}=0\;,\;\;\;\;
{\rm rot}\; ^{\alpha}{\bf H}=\gamma_{\alpha} 
\left({\bf E}_{\alpha}+\mu_{\alpha}\left[{\bf V}\times\;^{\alpha}{\bf 
H}\right]\right)
\;,\;\;\;\;\alpha={\sf in}, {\sf ex}\;,
\label{maxx1}
\end{eqnarray}
can be simplified considerably by applying Maxwell equations (\ref{maxx1}) 
for non--conducting media ($\gamma_{\sf in}=\gamma_{\sf ex}=0$) 
to Navier--Stokes equation (\ref{nav1}). $\gamma_{\sf in}$ and 
$\gamma_{\sf ex}$ denote conductivities of the jet's interior and exterior, 
respectively, $T_{jk}$ stands for magnetic stress tensor, and ${\bf 
V}^{\sf in}$ is local fluid velocity. Finally the magnetohydrodynamic 
problem is decoupled into the hydrodynamic and magnetostatic parts :
\begin{eqnarray} 
{\rm div}{\bf V}^{\sf in}=0\;,\;\;\;\;\frac{\partial 
{\bf V}^{\sf in}}{\partial t}=-\frac{1}{\rho}{\rm grad}\;P^{\sf in}\;,
\;\;\;\;\mbox{and}\;\;\;\;{\rm div}\; ^{\alpha}{\bf H}=0\;,\;\;\;\;
{\rm rot}\; ^{\alpha}{\bf H}=0\;,\;\;\;\alpha={\sf in}, {\sf ex}\;.
\label{maxx2}
\end{eqnarray}
A standard way to solve the boundary problem (\ref{maxx2}) is to introduce 
the Stokes stream function $\Psi(r,z,t)$ and magnetic potentials 
$\Phi_{\sf in}(r,z,t),\Phi_{\sf ex}(r,z,t)$
\begin{eqnarray}
V_r=-\frac{1}{r}\frac{\partial \Psi}{\partial z}\;,\;\;\;
V_z=\frac{1}{r}\frac{\partial \Psi}{\partial r}\;,\;\;\;
^{\alpha}{\bf H}={\bf H}_0+^{\alpha}{\bf H}^1\;,\;\;\;
^{\alpha}{\bf H}^1=-\nabla\Phi_{\alpha}\;,\;\;\;\alpha={\sf in}, {\sf ex}\;.
\label{maxx3}
\end{eqnarray}
This gives the following governing equations
\begin{eqnarray}
\frac{1}{\rho}\frac{\partial P^{\sf in}}{\partial z}+
\frac{1}{r}\frac{\partial^2 \Psi}{\partial r\partial t}=0\;,\;\;\;\;
\frac{1}{\rho}\frac{\partial P^{\sf in}}{\partial r}-
\frac{1}{r}\frac{\partial^2 \Psi}{\partial z\partial t}=0\;,\;\;\;\;
\left(\Delta_{2} + \frac{\partial^2 }{\partial 
z^2}\right)\Phi_{\alpha}=0\;,\;\;\alpha={\sf in}, {\sf ex}\;.
\label{maxx4}
\end{eqnarray}
These equations must be supplemented by four boundary conditions 
which are derived in the next Section.
\subsection{Boundary conditions}
\label{boundar1}
It is neseccary to apply boundary conditions (\ref{max2}), which are imposed 
on {\bf H} and {\bf B}, as well as those for the hydrodynamic variables. First, 
the velocity $V_r$ must be compatible, at $r=R$, with the assumed form of the 
deformed boundary $\partial \zeta/\partial t$. Second, at the free surface of 
a liquid jet the jump in stress must be balanced by Laplace pressure 
\cite{land84}, 
\begin{eqnarray}
\left[T_{rz}\right]^{\sf in}_{\sf ex} e_z+\left[T_{rr}\right]^{\sf 
in}_{\sf ex} e_r=2\sigma {\cal H}e_r\;,\;\;\;
\left[T_{zz}\right]^{\sf in}_{\sf ex} 
e_z+\left[T_{zr}\right]^{\sf in}_{\sf ex} e_r=2\sigma {\cal H}e_z\;,
\label{bonstr2}
\end{eqnarray}
where $\left[T_{jk}\right]^{\sf in}_{\sf ex}=T_{jk}^{\sf in}-
T_{jk}^{\sf ex}$ and ${\cal H}$ is mean surface curvature decomposed as 
\begin{eqnarray}
{\cal H}={\cal H}_0+{\cal H}_1\;,\;\;\;{\cal H}_0=\frac{1}{2R}\;,\;\;\;
{\cal H}_1=-\frac{1}{2}\left(\frac{\zeta}{R^2}+\frac{\partial^2 \zeta}
{\partial z^2}\right)\propto \epsilon=\frac{\zeta_0}{R}\;.
\label{lapla1}
\end{eqnarray}
By virtue of (\ref{max2a}) we get
\begin{eqnarray}
\left[T_{rr}\right]^{\sf in}_{\sf ex} -2\sigma {\cal H}=
\left[T_{rz}\right]^{\sf in}_{\sf ex} \frac{\partial \zeta}{\partial 
z}\;,\;\;\;
\left[T_{zr}\right]^{\sf in}_{\sf ex}=
\left(\left[T_{zz}\right]^{\sf in}_{\sf ex}-2\sigma {\cal H}\right)
\frac{\partial \zeta}{\partial z}\;,\;\;\;\mbox{or}\nonumber\\
\left[T_{rr}\right]^{\sf in}_{\sf ex} -2\sigma {\cal H}=
\left(\left[T_{zz}\right]^{\sf in}_{\sf ex}-2\sigma {\cal H}\right)
\left(\frac{\partial \zeta}{\partial z}\right)^2\;,\;\;\;
\left[T_{zr}\right]^{\sf in}_{\sf ex}=
\left(\left[T_{zz}\right]^{\sf in}_{\sf ex}-2\sigma {\cal H}\right)
\frac{\partial \zeta}{\partial z}\;,
\label{bonstr3}
\end{eqnarray}
where
\begin{eqnarray}
T_{rr}^{\sf in}=P^{\sf in}+
\frac{\mu}{2}\left\{(\;^{\sf in}H_z)^2-(\;^{\sf in}H_r)^2\right\}\;,\;\;\;
T_{zz}^{\sf in}=P^{\sf in}+
\frac{\mu}{2}\left\{(\;^{\sf in}H_r)^2-(\;^{\sf in}H_z)^2\right\}\;,\;\;\;
T_{zr}^{\sf in}=-\mu\; ^{\sf in}H_r\; ^{\sf in}H_z\;,\nonumber\\
T_{rr}^{\sf ex}=\frac{\mu_0}{2}\left\{(\;^{\sf ex}H_z)^2-(\;^{\sf 
ex}H_r)^2\right\}\;,\;\;\;
T_{zz}^{\sf ex}=\frac{\mu_0}{2}\left\{(\;^{\sf ex}H_r)^2-(\;^{\sf
ex}H_z)^2\right\}\;,\;\;\;
T_{zr}^{\sf ex}=-\mu_0\; ^{\sf ex}H_r\; ^{\sf ex}H_z\;.
\nonumber
\end{eqnarray}
Recalling that $\partial \zeta/\partial z \propto \epsilon$ and 
\begin{eqnarray}
^{\sf in}H_z-H_0=\;^{\sf in}H_z^1\propto \epsilon\;,\;\;
^{\sf ex}H_z-H_0=\;^{\sf ex}H_z^1\propto \epsilon\;,\;\;
^{\sf in}H_r=\;^{\sf in}H_r^1\propto \epsilon\;,\;\;
^{\sf ex}H_r=\;^{\sf ex}H_r^1\propto \epsilon\;,\nonumber\\
(^{\sf in}H_z)^2=H_0^2+2H_0\;^{\sf in}H_z^1+{\cal O}(\epsilon^2)\;,\;\;
(^{\sf ex}H_z)^2=H_0^2+2H_0\;^{\sf ex}H_z^1+{\cal O}(\epsilon^2)\;,\;\;
(^{\sf in}H_r)^2=(^{\sf ex}H_r)^2= {\cal O}(\epsilon^2)\nonumber
\end{eqnarray}
we get within the 1st order approximation in $\epsilon$
\begin{eqnarray}
\left[T_{rr}\right]^{\sf in}_{\sf ex}-2\sigma {\cal H}&=&P^{\sf in}
-2\sigma {\cal H}+
\left\{\frac{\mu}{2}(\;^{\sf in}H_z)^2-
\frac{\mu_0}{2}(\;^{\sf ex}H_z)^2\right\}+
\left\{\frac{\mu_0}{2}(\;^{\sf ex}H_r)^2-
\frac{\mu}{2}(\;^{\sf in}H_r)^2\right\}=\nonumber\\
&&P^{\sf in}-2\sigma {\cal H}+\frac{\mu_0\chi}{2}H_0^2+
H_0\left(\mu\;^{\sf in}H_z^1-\mu_0\;^{\sf ex}H_z^1\right)+
{\cal O}(\epsilon^2)\;,\label{rr1}\\
\left[T_{zz}\right]^{\sf in}_{\sf ex}-2\sigma {\cal H}&=&P^{\sf in}
-2\sigma {\cal H}+
\left\{\frac{\mu}{2}(\;^{\sf in}H_r)^2-\frac{\mu_0}{2}(\;^{\sf 
ex}H_r)^2\right\}+
\left\{\frac{\mu_0}{2}(\;^{\sf ex}H_z)^2-\frac{\mu}{2}(\;^{\sf 
in}H_z)^2\right\}=
\nonumber\\
&&P^{\sf in}-2\sigma {\cal H}-\frac{\mu_0\chi}{2}H_0^2-
H_0\left(\mu\;^{\sf in}H_z^1-\mu_0\;^{\sf ex}H_z^1\right)+
{\cal O}(\epsilon^2)\;.
\label{zz1}\\
\left[T_{zr}\right]^{\sf in}_{\sf ex}&=&\mu_0\; ^{\sf ex}H_r\; ^{\sf ex}H_z-
\mu\; ^{\sf in}H_r\; ^{\sf in}H_z=H_0 \left(\mu_0\; ^{\sf ex}H_r^1-
\mu\; ^{\sf in}H_r^1\right)+{\cal O}(\epsilon^2)\;.
\label{zr1}
\end{eqnarray}
Assuming $P^{\sf in}_1\propto \epsilon$ and combining the first 
equation in (\ref{bonstr3}) with (\ref{rr1}) we obtain
\begin{eqnarray}
P^{\sf in}-2\sigma {\cal H}+\frac{\mu_0\chi}{2}H_0^2+ H_0
\left(\mu\;^{\sf in}H_z^1-\mu_0\;^{\sf ex}H_z^1\right)=0\;\;\rightarrow\;
\left\{ \begin{array}{l}
P^{\sf in}_0-2\sigma {\cal H}_0+\frac{1}{2}\mu_0\chi H_0^2=0\;,\\
P^{\sf in}_1-2\sigma {\cal H}_1+ H_0
\left(\mu\;^{\sf in}H_z^1-\mu_0\;^{\sf ex}H_z^1\right)=0\;.
\end{array}\right.
\nonumber
\end{eqnarray}
This gives the unperturbed pressure within the cylindrical jet as,
\begin{eqnarray}
P^{\sf in}_0=\frac{\sigma}{R}-\chi\frac{\mu_0 H_0^2}{2}\;.
\label{bonstr5}
\end{eqnarray}
Combining the second boundary condition in (\ref{bonstr3}) with (\ref{zz1}), 
gives 
\begin{eqnarray}
\left(\mu_0\; ^{\sf ex}H_r-\mu\; ^{\sf in}H_r\right)H_0=
\left(P^{\sf in}_0-2\sigma {\cal H}_0-\frac{\mu_0\chi}{2}H_0^2\right)
\frac{\partial \zeta}{\partial z}\;.
\label{bonstr7} 
\end{eqnarray}
Both equations (\ref{bonstr5}) and (\ref{bonstr7}) lead to the conclusion 
\begin{eqnarray}
\mu_0\; ^{\sf ex}H_r-\mu\; ^{\sf in}H_r=
-\mu_0\chi H_0 \frac{\partial \zeta}{\partial z}\;,
\label{bonstr8}
\end{eqnarray}
that coincides with the second static boundary conditions in 
(\ref{max2}). Thus we arrive at 
\begin{eqnarray}
&&P^{\sf in}_1+\sigma\left(\frac{\zeta}{R^2}+\frac{\partial^2
\zeta}{\partial z^2}\right)+ H_0
\left(\mu\;^{\sf in}H_z^1-\mu_0\;^{\sf ex}H_z^1\right)=0\;,\;\;\;\;
V_r=\frac{\partial \zeta}{\partial t}\;,\label{bonstr9}\\
&&(1+\chi)\; ^{\sf in}H_r-\;^{\sf ex}H_r=
\chi H_0 \frac{\partial \zeta}{\partial z}\;,\;\;\;\;
^{\sf in}H_z^1=\;^{\sf ex}H_z^1\;,
\label{bonstr10} 
\end{eqnarray}
or using the notations of (\ref{maxx3})
\begin{eqnarray}
&&P^{\sf in}_1+\sigma\left(\frac{\zeta}{R^2}+\frac{\partial^2
\zeta}{\partial z^2}\right)+ H_0
\left(\mu_0\frac{\partial \Phi_{\sf ex}}{\partial z}-
\mu\frac{\partial \Phi_{\sf in}}{\partial z}\right)=0\;,\;\;\;\;\;\;
\frac{1}{r}\frac{\partial \Psi}{\partial z}+
\frac{\partial \zeta}{\partial t}=0\;,\label{bonstr9a}\\
&&\frac{\partial \Phi_{\sf ex}}{\partial r}-\left(1+\chi\right)
\frac{\partial \Phi_{\sf in}}{\partial r}=
\chi H\frac{\partial \zeta}{\partial z}\;,
\;\;\;\;\;\;\frac{\partial \Phi_{\sf ex}}{\partial z}=
\frac{\partial \Phi_{\sf in}}{\partial z}\;.
\label{bonstr10a}
\end{eqnarray}
Assuming that an axisymmetrical disturbance, characterized by a 
wavelength $2\pi/k$, increases exponentially in time with the growth rate 
$s$, gives,
\begin{eqnarray}
\{\Phi_{\sf in},\;\Phi_{\sf ex},\;\Psi,\;\zeta,\;P^{\sf in}_1\}=
\{i\;\phi_{\sf in}(r),\;i\;\phi_{\sf ex}(r),\;i\;\psi(r),\;\varsigma(r),\;p(r)\}
\times e^{st+ikz}\;.
\label{disper4}
\end{eqnarray}
Consequently, the following boundary conditions hold
\begin{eqnarray}
&&p+\sigma\left(\frac{1}{R^2}-k^2\right)\varsigma+ kH_0\left(\mu\phi_{\sf ex}-
\mu_0\phi_{\sf in}\right)=0\;,\;\;\;\;
s\varsigma=k\frac{\psi}{r}\;,\label{disper5}\\
&&\frac{\partial \phi_{\sf ex}}{\partial r}-\left(1+\chi\right)
\frac{\partial \phi_{\sf in}}{\partial r}=k\chi H_0 \varsigma\;,\;\;\;
\phi_{\sf ex}=\phi_{\sf in}\;.
\nonumber
\end{eqnarray}
The presence of the field--dependent term, in the first equation of 
(\ref{disper5}), indicates that the coupling of hydrodynamics and 
magnetostatics in the boundary conditions must break the Rayleigh 
dispersion relation (\ref{introd01}) for non--conducting jet.
\subsection{Dispersion relation}
\label{disp1}
\begin{figure}[h]
\vspace{.5cm}
\centerline{\psfig{figure=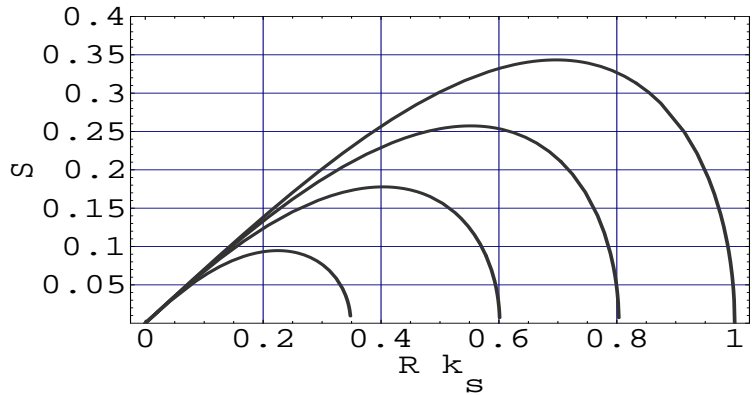,height=5.5cm,width=10cm}}
\vspace{-.5cm}
\caption{A plot of the dispersion relation $s(\varpi_s)$ for positive
susceptibilities $\chi=0, 1, 3, 10$, from right to left.}
\label{figplat1a}
\end{figure}

Inserting (\ref{disper4}) into (\ref{maxx4}) results in the following 
amplitude equations,
\begin{eqnarray}
\frac{1}{k}\frac{\partial p}{\partial r}+s\rho\frac{\psi}{r}=0\;,\;\;\;
kp+s\rho\frac{1}{r}\frac{\partial \psi}{\partial r}=0\;,\;\;\;
(\Delta_{2c}-k^2)\phi_{\sf in}=0\;,\;\;\;
(\Delta_{2c}-k^2)\phi_{\sf ex}=0\;.
\label{disper6}
\end{eqnarray}
The latter have fundamental solutions that are finite at $r=0$ and 
$r=\infty$, as follows
\begin{eqnarray}
\psi(r)=A_1krI_1(kr)\;,\;\;\;p(r)=-A_1s\rho kI_0(kr)\;,\;\;\;
\phi_{{\sf in}}(r)=A_2I_0(kr)\;,\;\;\;\phi_{{\sf ex}}(r)=A_3K_0(kr)\;,
\label{disper7}
\end{eqnarray}
where $A_i$ are three indeterminate coefficients. Substituting (\ref{disper7}) 
into (\ref{disper5}) we get the following dispersion relation
\begin{eqnarray}   
s^2(\varpi)=\frac{\sigma}{R^3 \rho} \frac{\varpi
I_1(\varpi)}{I_0(\varpi)}(1-\varpi^2)-\chi^2\varpi^3\frac{\mu_0H_0^2}{R^2 \rho}
\frac{I_1(\varpi)K_0(\varpi)}{T(\varpi,\chi)}\;.
\label{disper8}
\end{eqnarray}
(see Figure \ref{figplat1a}). The last expression clearly differs from 
Chandrasekhar's result (\ref{introd01}). Moreover, (\ref{disper8}) can be 
recasted as  
\begin{eqnarray}
s_{nc}^2(\varpi)=-\frac{\sigma}{R^3 \rho} \frac{\varpi 
I_1(\varpi)}{I_0(\varpi)} f(\varpi,\chi,H_0)\;,
\label{disper9}
\end{eqnarray}
where $f(\varpi,\chi,H_0)$ can be recognized as the dimensionless excess 
free energy found in (\ref{finl1}). Expression (\ref{disper9}) confirms 
the equivalence (\ref{intro10}) between the static and dynamic approaches 
in the problem of cut--off wavelength since both non--trivial zeroes, 
$\varpi_d$ and $\varpi_s$, of $s_{nc}(\varpi_d)$ and $f(\varpi_s,\chi,H_0)$, 
respectively, coincide.

\noindent
It is noteworthy that in the limit $\chi\to \infty$,
\begin{eqnarray}
s_{nc}^2(\varpi)\simeq \frac{\sigma}{R^3 \rho}
\left\{\frac{\varpi I_1(\varpi)}{I_0(\varpi)}(1-\varpi^2)-
\varpi^2\left(\frac{H}{H_{Ch}}\right)^2\right\}\;,
\label{disper10}
\end{eqnarray}
which coincides with the dispersion relation (\ref{introd01}) for a 
superconducting jet in the longwave limit $\varpi\to 0$.

\section{Conclusion}
\label{Conc}
We have revised the theory of capillary instability of isothermal 
incompressible inviscid and non--conducting liquid jets in a uniform 
magnetic field which was treated by Chandrasekhar \cite{Chan61} about 50 
years ago. The main result of \cite{Chan61} states that for non--conducting 
jets the field has no effect on their stability and the dispersion relation 
is given by Rayleigh's theory for free jets. This statement contradicts 
the free energy approach (Plateau problem) in the framework of static 
instability of a magnetazible liquid cylinder in a uniform field. 

We have found a new dispersion relation for magnetocapillary instability 
in such jets. The relation differs from that given by Chandrasekhar and 
agrees with the result for the cut--off wavelength obtained from the static 
approach. This reflects a deep equivalence between the static approach 
(the threshold of static instability which is concerned with an excess 
free energy) and the dynamic approach (the bifurcation of the first 
non--trivial steady state of the inviscid hydrodynamic system). In this 
context, the existence of critical magnetic field which stabilizes jets 
with finite susceptibility is established. It is shown that the jet is
stabilized by the field irrespective of its being para-- or diamagnetic,
but the extent of stabilization is different.

\newpage
\appendix
\renewcommand{\theequation}{\thesection\arabic{equation}}
\section{Contribution of the magnetic field to free energy}
\label{appendix1}
\setcounter{equation}{0}
Evaluate the contribution $\mu_0 U/2$ of the magnetic field inside 
$\Omega_{cyl}$ and outside ${\mathbb R}^3\setminus \Omega_{cyl}$ of the 
disturbed liquid cylinder to the excess free energy $W$ (see equation 
(\ref{stat8a}))
\begin{eqnarray}
U&=&\chi\left[\int_{\Omega_{cyl}}\; ^{\sf in}{\bf H}^2({\bf r})dv-
\int_{\Omega_{cyl}^0}H_0^2dv \right]+
\int_{\Omega_{cyl}}\; ^{\sf in}{\bf H}^2({\bf r})dv+
\int_{{\mathbb R}^3\setminus \Omega_{cyl}}\;^{\sf ex}{\bf H}^2({\bf r})dv-
\int_{{\mathbb R}^3}H_0^2dv\nonumber\\
&=&\chi\left[ 
\int_{\Omega_{cyl}}\left\{\;\left(H_0+^{\sf in}\!\!H_z^1\right)^2+
\left(^{\sf in}H_r^1\right)^2\right\}dv-\int_{\Omega_{cyl}^0}H_0^2dv\right]+
\nonumber\\
&&
\int_{\Omega_{cyl}}\;\left\{\left(H_0+^{\sf in}\!\!H_z^1\right)^2+
\left(^{\sf in}H_r^1\right)^2\right\}dv+\int_{{\mathbb R}^3\setminus 
\Omega_{cyl}}\;
\left\{\left(H_0+^{\sf ex}\!\!H_z^1\right)^2+
\left(^{\sf ex}H_r^1\right)^2\right\}dv-\int_{{\mathbb R}^3}H_0^2dv\nonumber\\
&=&\chi\left[
H_0^2\left(\int_{\Omega_{cyl}}dv-\int_{\Omega_{cyl}^0}dv\right)+
\int_{\Omega_{cyl}}\left\{2H_0\;^{\sf in}H_z^1+\left(^{\sf in}H_z^1\right)^2+
\left(^{\sf in}H_r^1\right)^2\right\}dv\right]+\nonumber\\
&&\int_{\Omega_{cyl}}\left\{2H_0\;^{\sf in}H_z^1+
\left(^{\sf in}H_z^1\right)^2+\left(^{\sf in}H_r^1\right)^2\right\}dv+
\int_{{\mathbb R}^3\setminus \Omega_{cyl}}
\left\{2H_0\;^{\sf ex}H_z^1+
\left(^{\sf ex}H_z^1\right)^2+\left(^{\sf ex}H_r^1\right)^2\right\}dv  
\nonumber\\
&=&(1+\chi)\int_{\Omega_{cyl}}\left\{2H_0\;^{\sf in}H_z^1+
\left(^{\sf in}H_z^1\right)^2+\left(^{\sf in}H_r^1\right)^2\right\}dv+
\int_{{\mathbb R}^3\setminus \Omega_{cyl}}
\left\{2H_0\;^{\sf ex}H_z^1+
\left(^{\sf ex}H_z^1\right)^2+\left(^{\sf ex}H_r^1\right)^2\right\}dv
\nonumber\\
&=&(1+\chi)\int_{\Omega_{cyl}}
\left\{\;\left(^{\sf in}H_z^1\right)^2+
\left(^{\sf in}H_r^1\right)^2\right\}dv+
\int_{{\mathbb R}^3\setminus \Omega_{cyl}}
\left\{\;\left(^{\sf ex}H_z^1\right)^2+
\left(^{\sf ex}H_r^1\right)^2\right\}dv+
\nonumber\\
&&2H_0\left((1+\chi)\int_{\Omega_{cyl}}\;^{\sf in}H_z^1dv+
\int_{{\mathbb R}^3\setminus \Omega_{cyl}}\;^{\sf ex}H_z^1dv\right)\;.
\nonumber
\end{eqnarray}
Hence, we conclude that
\begin{eqnarray}
U=\left(C_1\epsilon^2\chi^2+2C_2\epsilon\chi\right)H_0^2\;,
\label{stat5a}
\end{eqnarray}
where
\begin{eqnarray}
\frac{1}{\pi L}C_1&=&\frac{1+\chi}{\pi L}\int_{\Omega_{cyl}^0}
\left\{\;\left(^{\sf in}h_z^1\right)^2+
\left(^{\sf in}h_r^1\right)^2\right\}dv+
\frac{1}{\pi L}\int_{{\mathbb R}^3\setminus \Omega_{cyl}^0}
\left\{\;\left(^{\sf ex}h_z^1\right)^2+
\left(^{\sf ex}h_r^1\right)^2\right\}dv\;,
\label{stat5b}\\
\frac{1}{\pi L}C_2&=&\frac{1+\chi}{\pi L}\int_{\Omega_{cyl}}\;^{\sf in}h_z^1dv+
\frac{1}{\pi L}\int_{{\mathbb R}^3\setminus \Omega_{cyl}}\;^{\sf ex}h_z^1dv\;.
\label{stat5ba}
\end{eqnarray}
Recall the expression (\ref{stat8a}) for $W$
\begin{eqnarray}
W&=&\frac{\pi}{2}\left(\varpi^2-1\right)\sigma \epsilon^2RL-
\left(C_1\epsilon^2\chi^2+2C_2\epsilon\chi\right)\frac{\mu_0H_0^2}{2}=
\nonumber\\
&&\frac{\pi L\sigma R}{2}\left[\epsilon^2\left(\varpi^2-1\right)-
\left(\epsilon^2\frac{C_1}{\pi L}\chi+
2\epsilon\frac{C_2}{\pi L}\right)\frac{\chi\mu_0H_0^2}{\sigma R}\right]\;.
\end{eqnarray}
Making use of the integration
\begin{eqnarray}
&&\int_0^{u}\left[I_0^2(v)+I_1^2(v)\right]vdv=uI_0(u)I_1(u)\;,\;\;\;\;
\int_u^{\infty}\left[K_0^2(v)+K_1^2(v)\right]vdv=uK_0(u)K_1(u)\;,\nonumber\\
&&\int_0^{u}I_0(v)vdv=uI_1(u)\;,\;\;\;\;
\int_u^{\infty}K_0(v)vdv=uK_1(u)\;,\nonumber
\end{eqnarray}
and inserting (\ref{max8}) into (\ref{stat5b}) and (\ref{stat5ba}) gives,
\begin{eqnarray}
\frac{1}{\pi L}\int_{\Omega_{cyl}^0}
\left\{\;\left(^{\sf in}h_z^1\right)^2+
\left(^{\sf in}h_r^1\right)^2\right\}dv&=&
\frac{b^2}{\pi L}\cdot \frac{L}{2}\cdot 2\pi \frac{R^2}{\varpi^2}  
\int_0^{kR}\left[I_0^2(kr)+I_1^2(kr)\right](kr)d(kr)\nonumber\\
&=&\frac{b^2 R^2}{\varpi^2}\varpi I_0(\varpi) I_1(\varpi)=
\varpi^3 R^2\left(\frac{I_0(\varpi)K_0(\varpi)}
{T(\varpi)}\right)^2\frac{I_1(\varpi)}{I_0(\varpi)}
\nonumber\\
\frac{1}{\pi L}\int_{{\mathbb R}^3\setminus \Omega_{cyl}^0}
\left\{\;\left(^{\sf ex}h_z^1\right)^2+
\left(^{\sf ex}h_r^1\right)^2\right\}dv&=&
\frac{c^2}{\pi L}\cdot \frac{L}{2}\cdot 2\pi \frac{R^2}{\varpi^2}  
\int_{kR}^{\infty}\left[K_0^2(kr)+K_1^2(kr)\right](kr)d(kr)\nonumber\\
&=&\frac{c^2 R^2}{\varpi^2}\varpi K_0(\varpi) K_1(\varpi)=
\varpi^3 R^2\left(\frac{I_0(\varpi)K_0(\varpi)}
{T(\varpi)}\right)^2\frac{K_1(\varpi)}{K_0(\varpi)}\nonumber\\  
\frac{1}{\pi L}\int_{\Omega_{cyl}}\;^{\sf in}h_z^1dv&=&
-\frac{b}{\pi L}\cdot 2\pi\frac{R^2}{\varpi^2}
\int_0^{L}dz\cos kz\int_0^{kr(z)}I_0(kr)(kr)d(kr)\nonumber\\
&=&-\frac{2b}{L}\frac{R}{\varpi}\int_0^{L}\cos kz\cdot (R+\zeta_0\cos kz)
\cdot I_1(kR+k\zeta_0\cos kz)\;dz\nonumber\\
\frac{1}{\pi L}\int_{{\mathbb R}^3\setminus  \Omega_{cyl}}\;^{\sf ex}h_z^1dv&=&
-\frac{c}{\pi L}\cdot 2\pi \frac{R^2}{\varpi^2}
\int_0^{L}dz\cos kz\int_{kr(z)}^{\infty}K_0(kr)(kr)d(kr)\nonumber\\
&=&-\frac{2c}{L}\frac{R}{\varpi}\int_0^{L}\cos kz\cdot (R+\zeta_0\cos 
kz)\cdot K_1(kR+k\zeta_0\cos kz)\;dz\;.
\nonumber
\end{eqnarray}
Expanding $I_1(x+\varepsilon)$ and $K_1(x+\varepsilon)$ in the vicinity of 
$x=0$, up to the first order in $\varepsilon$, gives
$$
I_1(x+\varepsilon)=I_1(x)+\left(I_0(x)+I_2(x)\right)\frac{\varepsilon}{2}
\;,\;\;\;
K_1(x+\varepsilon)=K_1(x)-\left(K_0(x)+K_2(x)\right)\frac{\varepsilon}{2}\;.
$$
Finally we arrive at
\begin{eqnarray}
\frac{1}{\pi L}\int_{\Omega_{cyl}}\;^{\sf in}h_z^1dv&=&
-\frac{2b}{L}\frac{R}{\varpi}\int_0^{L}dz \cos kz\cdot
\left\{RI_1(\varpi)+\zeta_0\cos kz \left[I_1(\varpi)+\varpi
\frac{I_0(\varpi)+I_2(\varpi)}{2}\right]\right\}\nonumber\\
&=&-\frac{2b}{L}\frac{R}{\varpi}\zeta_0
\left[I_1(\varpi)+\varpi \frac{I_0(\varpi)+I_2(\varpi)}{2}\right]
\int_0^{L}dz \cos^2kz\cdot=-\epsilon b(\varpi,\chi)R^2 I_0(\varpi)\;,\nonumber\\
\frac{1}{\pi L}\int_{{\mathbb R}^3\setminus \Omega_{cyl}}\;^{\sf ex}h_z^1dv&=&
-\frac{2c}{L}\frac{R}{\varpi}\int_0^{L}dz \cos kz\cdot
\left\{RK_1(\varpi)+\zeta_0\cos kz \left[K_1(\varpi)-\varpi
\frac{K_0(\varpi)+K_2(\varpi)}{2}\right]\right\}=\nonumber\\
&=&-\frac{2c}{L}\frac{R}{\varpi}\zeta_0
\left[K_1(\varpi)-\varpi \frac{K_0(\varpi)+K_2(\varpi)}{2}\right]
\int_0^{L}dz \cos^2kz=\epsilon c(\varpi,\chi) R^2 K_0(\varpi)\;.
\nonumber
\end{eqnarray}
Thus, we get
\begin{eqnarray}
\frac{1}{\pi L}C_1&=&\varpi^3 R^2\left(\frac{
I_0(\varpi)K_0(\varpi)}{T(\varpi,\chi)}\right)^2
\left\{(1+\chi)\frac{I_1(\varpi)}{I_0(\varpi)}+
\frac{K_1(\varpi)}{K_0(\varpi)}\right\}=\varpi^2 R^2 
\frac{I_0(\varpi)K_0(\varpi)}{T(\varpi,\chi)}\;,\label{prelf1}\\   
\frac{1}{\pi L}C_2&=&-\epsilon \varpi^2 R^2\left[
(1+\chi)\frac{I_0(\varpi)K_0(\varpi)}{T(\varpi,\chi)}-
\frac{I_0(\varpi)K_0(\varpi)}{T(\varpi,\chi)}\right]=
-\epsilon \chi \varpi^2 R^2 
\frac{I_0(\varpi)K_0(\varpi)}{T(\varpi,\chi)}\;,
\label{prelf2}
\end{eqnarray}
that consequently gives,
\begin{eqnarray}
W=\frac{\pi L\sigma R}{2}\epsilon^2 \cdot f(\varpi,\chi)\;,\;\;\;
f(\varpi,\chi)=\varpi^2-1-\left(\frac{C_1}{\pi L}\chi+
2\frac{C_2}{\epsilon\pi L}\right)\frac{\chi\mu_0H_0^2}{\sigma R}\;.
\label{prelf3}  
\end{eqnarray}
Insertion (\ref{prelf1}) and (\ref{prelf2}) in the latter expression, 
we obtain
\begin{eqnarray}
f(\varpi,\chi)&=&\varpi^2-1+\chi^2\varpi^2\frac{\mu_0 R H_0^2}{\sigma}
\frac{I_0(\varpi)K_0(\varpi)}{T(\varpi,\chi)}\;\label{fin1}\;.
\label{prelf4}
\end{eqnarray}
\section{Asymptotics of expressions}
\label{appendix2}
\setcounter{equation}{0}
Consider the expressions obtained in Appendix \ref{appendix1} and evaluate 
its asymptotics. 
\begin{itemize}
\item $\varpi\to 0\;,\;\;|\chi|\ll 1$.
\end{itemize}
\begin{eqnarray}
I_1(\varpi)\sim \frac{\varpi}{2}\left(1+\frac{\varpi^2}{8}\right)\;,\;\;\;
I_0(\varpi)\sim 1+ \frac{\varpi^2}{4}\;,\;\;\;
K_0(\varpi)\sim \beta(\varpi)\left(1 
+\frac{\varpi^2}{4}\right)\;,\;\;\;
\beta(\varpi)=-\ln\frac{\gamma\varpi}{2}\;,
\nonumber
\end{eqnarray}
where $\gamma=0.577216$ is Euler' constant. The corresponding asymptotics 
for the dimensionless excess free energy $f(\varpi,\chi)$ and the 
critical field $H_{cr}(\varpi,\chi)$ read
\begin{eqnarray}
f(\varpi,\chi)&=&  
\varpi^2-1+\chi\varpi^2\beta(\varpi)
\left(1-\frac{\chi}{2}\varpi^2\beta(\varpi)\right)
\left(\frac{H_0}{H_{Ch}}\right)^2\;,\label{ener1}\\
H_{cr}(\varpi,\chi)&=&\frac{H_{Ch}}{\varpi}\frac{1}{\sqrt{-\chi\ln\varpi}}
\left(1-\frac{\chi}{2}\varpi^2\ln\varpi\right)\;,
\label{magfi1}
\end{eqnarray}


\begin{itemize}
\item $\varpi\to 0\;,\;\;\chi\to \infty$.
\end{itemize}
\begin{eqnarray}
f(\varpi,\chi)\sim
\varpi^2-1+2\left(1+\frac{\varpi^2}{8}\right)
\left(\frac{H_0}{H_{Ch}}\right)^2\;\;\rightarrow\;\;
H_{cr}(\varpi,\chi)=\frac{H_{Ch}}{\sqrt{2}}
\left(1-\frac{9}{16}\varpi^2\right)\;.\label{w0xin}
\end{eqnarray}
\begin{itemize}
\item $\varpi\to 1$
\end{itemize}
\begin{eqnarray}
f(\varpi,\chi)\sim 
\varpi^2-1+\frac{\chi I_0(1)K_0(1)}{1+\chi I_1(1)K_0(1)}
\left(\frac{H_0}{H_{Ch}}\right)^2\;,\;\;\;\;
H_{cr}(\varpi,\chi)= B_1 H_{Ch}\sqrt{B_2+\frac{1}{\chi}}
\sqrt{1-\varpi^2}\;,
\label{w1xin}
\end{eqnarray}  
where $B_1=1/\sqrt{I_0(1)K_0(1)}\simeq 1.3697,\;B_2=I_1(1)K_0(1)\simeq 
0.2379$.


\end{document}